\documentclass[aps,prx,twocolumn,floats,showpacs,superscriptaddress,nofootinbib]{revtex4-1}

\usepackage{feynmp}
\usepackage{graphicx,epsfig}
\usepackage{times}
\usepackage{graphics,dcolumn,bm,float}
\usepackage{amssymb,amsmath,rotate,color}
\usepackage[title,titletoc,toc]{appendix}
\usepackage{graphicx}
\usepackage{wrapfig}

\usepackage{lipsum}
\usepackage{mwe}

\usepackage[pagebackref=false,colorlinks,linkcolor=blue,citecolor=blue,urlcolor=magenta]{hyperref}

\DeclareMathOperator\sgn{sgn}

\usepackage[mathlines]{lineno}
\usepackage{hyperref}
\usepackage{breqn}
\usepackage{bbold}
\usepackage{pgfplots}
\usepackage{float}
\usepackage{tkz-euclide}
\usepackage{braket}
\usepackage{physics}
\usepackage[caption=false]{subfig}
\usepackage[export]{adjustbox}
\usepackage{tikz}
\usetikzlibrary{through,calc}
\usetikzlibrary{positioning}

\newcommand{\be}{\begin{equation}}
\newcommand{\ee}{\end{equation}}
\newcommand{\ep}{\epsilon}

\newcommand{\bearr}{\begin{eqnarray}}
\newcommand{\eearr}{\end{eqnarray}}
\newcommand{\nn}{\nonumber}

\newcommand{\bsq}{{\boldsymbol{q}}}
\newcommand{\bsk}{{\boldsymbol{k}}}

\newcommand{\bs}{\boldsymbol}

\begin{document}
\preprint{}
\title{Polarization tensor for tilted Dirac fermion materials: Covariance in deformed Minkowski spacetime
}
\author{Z. Jalali-Mola}
\email{jalali@physics.sharif.edu}
\affiliation{
Department of Physics$,$ Sharif University of  Technology$,$ Tehran 11155-9161$,$ Iran
}

\author{S.A. Jafari}
\email{jafari@physics.sharif.edu}
\affiliation{
Department of Physics$,$ Sharif University of  Technology$,$ Tehran 11155-9161$,$ Iran
}
\affiliation{Center of excellence for Complex Systems and Condensed Matter (CSCM)$,$
Sharif University of Technology$,$Tehran 1458889694$,$ Iran}

\date{\today}

\begin{abstract}
The rich structure of solid state physics provides us with Dirac materials the effective theory of which enjoys the Lorentz symmetry. 
In non-symmorphic lattices, the Lorentz symmetry will be deformed in a way that the null energy-momentum vectors will correspond to
on-shell condition for tilted Dirac cone dispersion. In this sense, tilted Dirac/Weyl materials can be viewed as solid state systems where the effective spacetime is non-Minkowski. 
In this work, we show that the polarization tensor for tilted Dirac cone systems acquires a covariant from only when the spacetime is considered to be
an appropriate deformation of the Minkowski spacetime. As a unique consequence of the deformation of the geometry of the spacetime 
felt by the electrons in tilted Dirac cone materials, the Coulomb density-density interactions will generate corrections in both longitudinal
and transverse channels. Therefore the transverse photons also participate in mediating the Coulomb forces, implying emergent Amperean 
forces associated with the tilt of the spacetime. 
\end{abstract}



\maketitle
\narrowtext

{\em Introduction:---}
Dirac solids are materials in which the continuum limit of the effective Hamiltonian is the Dirac equation. From geometric 
perspective, the spacetime felt by the electronic degrees of freedom is the Minkowski spacetime. An important isometry of this spacetime
is the Lorentz transformation. The interest in Dirac materials is perhaps due to the relativistic
covariance of the electronic theory in these systems which has many interesting consequences~\cite{Baskaran2007,CastroNeto,Katsnelsonbook},
including but not limited to pseudoscalar superconductivity~\cite{Faraei2017} with no counterparts in non-Dirac materials. 

Breaking the Lorentz symmetry in solid state physics is not a surprise. Indeed the effective
theory in most solids is only Galilean invariant~\cite{SchakelBook}.
But in tilted Dirac cone systems the Lorentz symmetry is broken in a very interesting way~\cite{TohidBorophene,JafariEB}. 
Any non-zero amount of tilt which we parameterize by two parameters $\bs{\zeta}=(\zeta_x,\zeta_y)$ breaks the
Lorentz symmetry, but still a residual symmetry is left behind which turnes out to be a deformation of the Lorentz group~\cite{JafariEB}. The tilt deformation of the Dirac equation can be achieved in three dimensional materials~\cite{Soluyanov2015,Goerbig2016PRL, Yun2016, Armitage2018} 
as well as in two space dimensions~\cite{TohidBorophene,Tarun2017,SaharTilt1,SaharTilt2}. Two dimensional
examples include the organic material~\cite{Katayama2006,Suzumura2014} and recently proposed 2D allotrope of boron, namely $8Pmmn$ borophene~\cite{Zhou2014,Zabolotskiy2016,Lopez2016,Feng2017}. 
The nice property of 2D borophene is that the amount of tilt can be controlled by the perpendicular electric field~\cite{TohidBorophene}.  
There are also recent experimental realization of over-tilted Dirac cone in  transition metal dichalcogenide PtTe$_2$~\cite{Yan2107} and  PdTe$_2$ superconductor~\cite{Noh}. 

Geometric interpretation of the tilt deformation of the Dirac cone immediately leads to the fact that
the spacetime felt by electrons in tilted Dirac cone materials is {\em not} the Minkowski spacetime, 
but a deformation thereof~\cite{JafariEB,TohidBorophene,Foster2019}. In this work, by brute force
calculation of the polarization tensor, we show that the metric of the modified Minkowski space appears
in a nice and covariant way, provided instead of the Minkowski metric $\eta^{\mu\nu}$, appropriate metric 
$g^{\mu\nu}$ is used~\cite{JafariEB}. This signifies that with respect to the electromagnetic response, the electrons in tilted Dirac cone solid behave as if they live 
in a deformed Minkowski spacetime. This immediately implies that the plasmon oscillations of the electron density 
in the tilted Dirac electron liquids are accompanied by both longitudinal and transverse electric fields.

{\em tilted Dirac model:---}
The  general form of effective Hamiltonian near the Dirac point for one of the tilted Dirac cones is given by,
  \begin{equation}
  H(k)=\hbar \begin{pmatrix}  v_{x,t}k_x + v_{y,t}k_y &  v_x k_x-i v_y k_y\\    v_x k_x+i v_y k_y &  v_{x,t}k_x + v_{y,t}k_y \end{pmatrix},
    \label{matrixform}
    \end{equation}
where the off-diagonal elements represent the cone-like feature of the system characterized with Fermi velocity scales $v_x$ and $v_y$. 
The diagonal elements characterize by the tilt velocities $v_{x,t}$ and $v_{y,t}$ represent the tilting characteristic~\cite{SaharTilt1,SaharTilt2}. 
Let us assume that the ratio between the tilt parameters and the Fermi velocity $v_F=v_x=v_x$ is given by $v_{x,t}=\zeta_x v_F$ and $v_{y,t}=\zeta_y v_F$.
We set the units by $\hbar=v_F=1$. Therefore the above Hamiltonian becomes, 
 \begin{equation}
    H(k)= \begin{pmatrix}  \zeta_{x}k_x + \zeta_{y}k_y &   k_x-i  k_y\\    k_x+i  k_y &  \zeta_{x}k_x + \zeta_{y}k_y \end{pmatrix}=\bs\zeta.\bsk\sigma_0+\bsk.\bs\sigma,
      \label{new-matrixform}
  \end{equation}
the spectrum of which is,
\be
E_{s}(k)=  s k +\bs{\zeta}.\bs{k},~~~
 \ket{\bs{k},s}= \frac{1}{\sqrt{2}}\begin{pmatrix} 1 \\ s e^{i\theta_{k}} \end{pmatrix},
 \label{dispersion.eqn}
 \ee 
where $s=\pm$, $E_{+}$ and $E_{-}$ are conduction and valence bands respectively. The $\bs{\zeta}$ is a dimensionless parameter  which describes the strength of the
tilt in the system.
The effective Hamiltonian around the other valley can be obtained from the above equation by 
$\bs{k}\to -\bs{k}$ and $\bs\sigma \to \bs\sigma^\star$~\cite{Kobayashi2007,Goerbig2008}. This model reduces to the graphene (upright Dirac cone) in the limit $\bs{v}_t=0$.

Alternative way to think about the above dispersion relation is to assume that it describes the 
null spacetime vectors in $2+1$ dimensional spacetime, namely $g^{\mu\nu}k_\mu k_\nu=0$ where $k_\mu=(E,\bs{k})$
and the covariant components of $g_{\mu\nu}$ are defined by  Painelev\'e-Gulstrand metric~\cite{Volovik2018} $ds^2=-dt^2+(d\bs{r}^2-\bs{v}_tdt)^2$
which is explicitly given by,
\be
g_{\mu \nu}=
\begin{pmatrix}
-1+\zeta^2 & -\zeta_x & -\zeta_y \\
 -\zeta_x & 1 & 0 \\
 -\zeta_y & 0 & 1
\end{pmatrix},
\label{tmetric.eqn}
\ee
where $\zeta^2=\zeta_x^2+\zeta_y^2$ is the magnitude of parameter $\bs{\zeta}$.  In type I tilted Dirac cone we have $0\le \zeta\leq1$~\cite{Kobayashi2007,Zhou2014} 
and in type II tilted Dirac cone is  $\zeta>1$~\cite{Noh,Yan2107}.
In $3+1$ dimensional systems a border corresponding to $\zeta=1$ corresponds to a standard black-hole horizon~\cite{Ryder,Padmanabhan}.
In $2+1$ dimensional systems such a border will correspond to BTZ black-holes~\cite{BTZ}.  
The purpose of the present work is to show that the metric in 
Eq.~\eqref{tmetric.eqn} naturally appears in the electromagnetic response of the tilted Dirac fermions encoded in the polarization tensor. 

An essential feature of the above dispersion relation is its anisotropic dependence on the wave vector $\bsk$
which is due to the vector $\bs{\zeta}$ and makes the calculation of correlation functions different and more complicated compared to graphene, especially in extrinsic case~\cite{SaharTilt1}. In our previous work, we have calculated the density-density correlation function in tilted Dirac cone systems. 
The physical consequence of tilt $\bs{\zeta}$ for the doped Dirac cone is that it generates a kink in the plasmon dispersion~\cite{SaharTilt1,SaharTilt2}. 
In recent work, we have built on Eq.~\eqref{tmetric.eqn} that states the spacetime felt by the electrons satisfying the tilted Dirac equation
is a deformation of the Minkowski spacetime, and have obtained the algebraic structure of such a deformed Minkowski spacetime~\cite{JafariEB}. 
In the present work, we calculate the entire polarization tensor
$\Pi^{\mu\nu}(q)=-\tau\langle j^\mu(q)j^\nu(q) \rangle$~\cite{Peskin} by brute force. Then we will show that it acquires a covariant form
{\em only} in a spacetime given by metric~\eqref{tmetric.eqn}. We will then discuss the consequences of non-zero components $\bs{\zeta}$ that mix space and time in this metric.

{\em Polarization tensor:---}
Let us start the calculation of the polarization tensor for tilted Dirac fermions with tilt parameter $\bs{\zeta}$.
In this work we will be interested in $|\zeta|<1$. Since we are interested in small momentum (long wavelength) response,
we will confine ourselves to a single valley. 
In general
 current response function in Lehmann representation  is given by,
  \bearr
 && \tilde\Pi^{\mu \nu}(\boldsymbol{q},\omega)=  \frac{g_s  }{A} \times\nn\\&& \lim\limits_{\ep\rightarrow 0}  \sum_{k,s , s'=\pm} \frac{n_{k,s}-n_{k+q,s'}}{ \omega+E_{k,s}-E_{k+q,s'}+i\ep} \tilde F_{s, s'}^{\mu \nu} (\bsk, \bsk'),
 \label{LR.eqn}
  \eearr
where $A$ is area of system and  $g_s$ is the spin degeneracy. Here $\ep$ is defined as an infinitesimal positive constant and Fermi  distribution function defined by $n_{k,s}$ in which is a step function at zero temperature and $\bsk'=\bsk+\bsq$ with direction of $\bsk$ and $\bsq$ along $x$ axis is $\theta_\bsk$ and $\phi$, respectively. 
Note that the valleys are not degenerate anymore. 
The corresponding results for the other valley can be obtaind by $\bs{\zeta}\to -\bs{\zeta}$.
The form factor $\tilde F_{s, s'}^{\mu \nu} (\bsk, \bsk')$ is the expectation value of current element as,
\be
\tilde F_{s, s'}^{\mu \nu} (\bsk, \bsk')=\langle\bsk,s|\tilde j^\mu|\bsk',s'\rangle\langle\bsk',s'|\tilde j^\nu|\bsk,s\rangle,
\label{form factor.eqn}
\ee  
where $\mu,\nu=0,1,2$ as the spacetime is $2+1$ dimensional, and $\tilde j^0$ stands for charge density $\tilde{\rho}$ and $\tilde{j}^{1(2)}$ 
stands for the spatial components $\tilde{j}^{x(y)}$ of the current density which can be directly extracted from
Eq.~\eqref{new-matrixform} and have following representation,
 \bearr
  \tilde{\rho}= \begin{pmatrix} 1 &  0\\   0&  1\end{pmatrix},~~~
  \tilde{j}^x= \begin{pmatrix}  \zeta_{x} &  1 \\   1 &  \zeta_{x}\end{pmatrix},~~~
  \tilde{j}^y=\begin{pmatrix}  \zeta_{y} & -i  \\    i   & \zeta_{y} \end{pmatrix}.
    \label{jx,jy}
\eearr
The Lehmann representation in Eq.~\eqref{LR.eqn} can be simplified to
\bearr
&&\tilde\Pi^{\mu \nu}(\boldsymbol{q},\Omega)=  \frac{g_s  }{A} \nn\\&&
\lim\limits_{\ep\rightarrow 0}  \sum_{k,s , s'=\pm}  \tilde F^{\mu\nu}_{s, s'} (\bsk,\bsq)  \frac{n_{k,s}-n_{k+q,s'}}{( \Omega+s |\bsk|-s' |\bsk+\bsq|)+i\ep},
\label{Pai.eqn}
\eearr
where $\Omega=\omega-\bsq.\bs{\zeta}$. Indeed the new representation reduces to the polarization tensor for graphene when $\bs{\zeta}\to 0$.
In this limit, we have $\Omega \to \omega$ and the matrix elements $\tilde F^{\mu\nu}_{s, s'}$ reduce to the corresponding ones, namely,
$F^{\mu\nu}_{s, s'}$ of graphene. 

Although one can directly evaluate the integrals appearing in Eq.~\eqref{LR.eqn} which has been done in 
appendix~\ref{alternative.sec}, nevertheless, a quick and neat way of obtaining the same result is to
note that in the undoped Dirac cone in both tilted and tilt-less situations the Fermi surface will be
a point node. Therefore one can relate the matrix elements $\tilde F$ of tilted Dirac cone with $F$ of
upright Dirac cones. To do this, let us note that $3$-current  $\tilde j^\mu$ (of tilted Dirac cones) given in Eq.~\eqref{jx,jy} 
and $ j^\mu$ (of the tilt-less limit) are related by,
\be
\tilde{\rho}=\rho,~~~\tilde{j}^x=\zeta_x\hat{\rho}+j^x,~~~j^y=\zeta_y\hat{\rho}+j^y,
\label{tildehat.eqn}
\ee
where quantities without tilde correspond to $\bs{\zeta}=0$ limit, and those with tilde correspond to the tilted system. 
Already at this basic level, it can be seen that the spatial components $\tilde j^i$ of the $3$-vector in tilted Dirac
system is a mixture of both temporal ($\rho$) and spatial components ($j^i$) of the tilt-less system. 
Eq.~\eqref{tildehat.eqn} immediately implies the following set of relations between the matrix elements $\tilde F$ and $F$:
\bearr
  &&\tilde F^{00}_{s,s'}(\bsk,\bsk')=F^{00}_{s,s'}(\bsk,\bsq),\nn\\
  &&\tilde F^{ii}_{s,s'}(\bsk,\bsk')=F^{ii}_{s,s'}(\bsk,\bsq)+2\zeta_i F^{i0}_{s,s'}(\bsk,\bsq)+
  \zeta_i^2 F^{00}_{s,s'}(\bsk,\bsq),\nn\\
  &&\tilde F^{i0}_{s,s'}(\bsk,\bsk')=\tilde F^{0i}_{s,s'}(\bsk,\bsk')=\zeta_iF^{00}_{s,s'}(\bsk,\bsq)+ F^{i0}_{s,s'}(\bsk,\bsq),
  \nn\\
  &&\tilde F^{ij}_{s,s'}(\bsk,\bsk')=\zeta_i\zeta_jF^{00}_{s,s'}(\bsk,\bsq)+F^{ij}_{s,s'}(\bsk,\bsq)\nn\\
  &&+\zeta_i F^{j0}_{s,s'}(\bsk,\bsq)+\zeta_j F^{i0}_{s,s'}(\bsk,\bsq).
 \label{form factor-g.eqn}
\eearr
Here the graphene form factors are denoted by $F^{\mu\nu}_{s,s'}(\bsk,\bsq)$ which has the following representation,
\bearr
&&F^{00}_{s,s'}(\bsk,\bsq)=1/2(1+s s'\cos(\theta_\bsk-\theta_{\bsk'})),\nn\\
&&F^{11}_{s,s'}(\bsk,\bsq)=1/2(1+s s'\cos(\theta_\bsk+\theta_{\bsk'})),\nn\\
&&F^{10}_{s,s'}(\bsk,\bsq)=1/2( s \cos\theta_\bsk+s'\cos\theta_{\bsk'}).
\eearr
The rest of the elements can be derived from the above results  by a simple shift of the angular variables
$\theta_{\bs k}\rightarrow \theta_{\bs k}+\phi$ where $\phi$  is the polar angle of $\bs{q}$ with respect 
to $k_x$-axis. This amounts to $\cos\theta_\bsk \to \cos(\theta_\bsk+\phi) $ and so on,
\bearr
&&F^{22}_{s,s'}(\bsk,\bsq)=F^{11}_{s,s'}(\bsk,q,\phi-\pi/2),\nn\\
&&F^{12}_{s,s'}(\bsk,\bsq)=F^{21}_{s,s'}(\bsk,\bsq)=1/2F^{11}_{s,s'}(\bsk,q,\phi-\pi/4)\nn\\
   &&-1/2F^{11}_{s,s'}(\bsk,q,\phi+\pi/4),\nn\\
&&F^{20}_{s,s'}(\bsk,\bsq)=F^{02}_{s,s'}(\bsk,\bsq)=F^{11}_{s,s'}(\bsk,q,\phi-\pi/2).
\eearr
The above relations between the matrix elements of $\tilde F$ and $F$, induces a similar relation between the 
tensors $\tilde \Pi$ of Eq.~\eqref{Pai.eqn} and $\Pi$ of the upright Dirac cone. The only additional point is that
to obtain the correct form of denominator in Eq.~\eqref{Pai.eqn} one has to supplement it with substitution of $\omega$ by $\Omega=\omega-\bsq.\bs{\zeta}$.

{\em Covariant representation:---}
Eq.~\eqref{form factor-g.eqn} supplemented by $\omega\to\Omega$ gives us the entire matrix elements of 
the polarization tensor. It turns out that the polarization tensor for undoped graphene~\cite{Wunsch2006,Hwang2007} can be
represented in a compact covariant form~\cite{Peskin,ZeeQFTBook,Son2007}
\be
    \Pi^{\mu\nu}(\omega,q)=\pi(q)[q^2 \eta^{\mu \nu}-q^\mu q^\nu],~~~~~\pi(q)=\frac{-g_s}{16\sqrt{q^2}}
    \label{covgraphene.eqn}
\ee
where $\eta_{\mu\nu}={\rm diag}(-1,1,1)$ is the metric of the 2+1 dimensional Minkowski metric. This relation
holds in arbitrary dimensions~\cite{Peskin,ZeeQFTBook}, $q^2=q^\alpha q_\alpha=-\omega^2+\bsq^2$ is the length of the $3$-vector $q^\alpha$
defined with respect to the above metric, $q^\mu=\eta^{\mu\alpha}q_\alpha$. The function $\pi(q)$
is also an invariant (scalar) with respect to isometries of the Minkowski metric (i.e. with respect to rotations and Lorentz transformations). 
In fact, $\Pi^{\mu\nu}$ being a tensor has to be constructed from the only tensors available in the theory, namely $\eta^{\mu\nu}$ and $q^\mu q^\nu$
is severely restricted by the general covariance principle. 
Furthermore, imposing the Ward identity $q_\mu\Pi^{\mu\nu}=0$ fixes the
form of the terms in the square brackets. It only remains to specify the scalar $\pi(q)$ which can be
obtained from the graphene literature. 
In the above equation  the real (imaginary) part is nonzero when $q>|\omega|$ ($q<|\omega|$) and furthermore
$\Im\Pi^{\mu\nu}(\omega,q)$ is equal to $\sgn(\omega)\Im\Pi^{\mu\nu}$. 
We should notice that the graphene response function in addition to spin degeneracy has valley degeneracy which we ignore it in our calculation for tilted Dirac cone, 
hence $g_s$ in above equation points to the spin degeneracy. 

We are now ready to construct the explicit form of the polarization tensor $\tilde \Pi$ for tilted Dirac cone systems. The components are given by,
\bearr
&&\tilde\Pi^{00}(\boldsymbol{q},\Omega)=
  \{q^2 \} \tilde \pi(q),
  \nn\\
&&\tilde\Pi^{ii}(\boldsymbol{q},\Omega)=
  \{(\Omega^2-q^2)+q_i^2+q^2\zeta_i^2+2\Omega q_i \zeta_i
 \} \tilde \pi(q),
 \nn\\
  &&\tilde\Pi^{i0}(\boldsymbol{q},\Omega)=\tilde\Pi^{0i}(\boldsymbol{q},\Omega)=
    \{q_i \Omega+q^2\zeta_i  \}   \tilde\pi(q),
\nn\\
   &&\tilde\Pi^{ij}(\boldsymbol{q},\Omega)=
     \{q_i q_j+q^2\zeta_i \zeta_j+\Omega q_i \zeta_j+\Omega q_j\zeta_i
    \}  \tilde\pi(q),
\label{xx-yy-xy-undoped.eqn}
 \eearr
where as usual $i,j$ run over spatial indices $1,2$ and 
$\tilde\pi(q)=\pi(\bsq,\Omega)$ is the scalar defined in Eq.~\eqref{covgraphene.eqn} and $\Omega>0$. 
The only possible factor that can generate the imaginary part is the scalar part $\tilde \pi(q)$.
Therefore a purely real (imaginary) response is obtained when $q>|\Omega|$ ($q<|\Omega|$) and further 
$\Im\Pi^{\mu\nu}(\Omega,q)$is proportional to the $\sgn (\Omega)$. 
The cuttoff term that arises from the 
Keramerz-Kronig integration relation disappears by requiring the gauge invariance~\cite{principi,SaharStoner}. 
 
The set of equations~\eqref{xx-yy-xy-undoped.eqn} can be nicely summarized in the following
covariant expression,
 \be
 \tilde \Pi^{\mu \nu}= [q^2 g^{\mu \nu}-q^{\mu} q^{\nu}]  \tilde\pi({q}^2),~~~~~\tilde\pi({q})=\frac{-g_s}{16\sqrt{ q^2}},
 \label{tilted-cov.eqn}
 \ee
where $q_\mu=(-\omega,q_x,q_y)$ and and $q^\mu$ is naturally given by,
\be
  {q}^\mu= g^{\mu \nu} q_{\nu} \Rightarrow {q}^2\equiv q^\mu q_\mu=-\Omega^2+\bsq^2,
  \label{q2Om2.eqn}
\ee 
where $g^{\mu\nu}$ is the inverse of the metric~\eqref{tmetric.eqn} which is given by,
\be
g^{\mu \nu}=
\begin{pmatrix}
-1 &- \zeta_x & -\zeta_y \\
 -\zeta_x & 1-\zeta_x^2 & -\zeta_x \zeta_y \\
  -\zeta_y & -\zeta_x \zeta_y &  1-\zeta_y^2
\end{pmatrix}.
\label{cov.eqn}
\ee
Eq.~\eqref{q2Om2.eqn} identifies the combination $-\Omega^2+\bsq^2$ simply as an invariant in the
spacetime given by metric~\eqref{tmetric.eqn}. {\em The covariant from in Eq.~\eqref{cov.eqn} can not
be obtained with the metric $\eta^{\mu\nu}$.} Therefore the covariance of the polarization tensor for
tilted Dirac matter at zero density can only be attained when the metric~\eqref{tmetric.eqn} replaces
the Minkowski metric $\eta^{\mu\nu}$. Eq.~\eqref{cov.eqn} pleasantly satisfies the Ward identity 
$q^\mu\tilde\Pi^{\mu\nu}=0$. Needless to say that when the tile $\bs{\zeta}\to 0$, the metric $g^{\mu\nu}$
reduces to $\eta^{\mu\nu}$ of the standard Minkowski spacetime, and hence the electromagnetic response
$\tilde\Pi^{\mu\nu}$ trivially reduces to $\Pi^{\mu\nu}$. 

What is the difference between the geometry given by $g_{\mu\nu}$ and the standard Minkowski spacetime when the density-density Coulomb interaction in the materials is turned on? This is the subject of the next section.


 \begin{figure}[t]
     \centerline{\includegraphics[width = .5\textwidth] {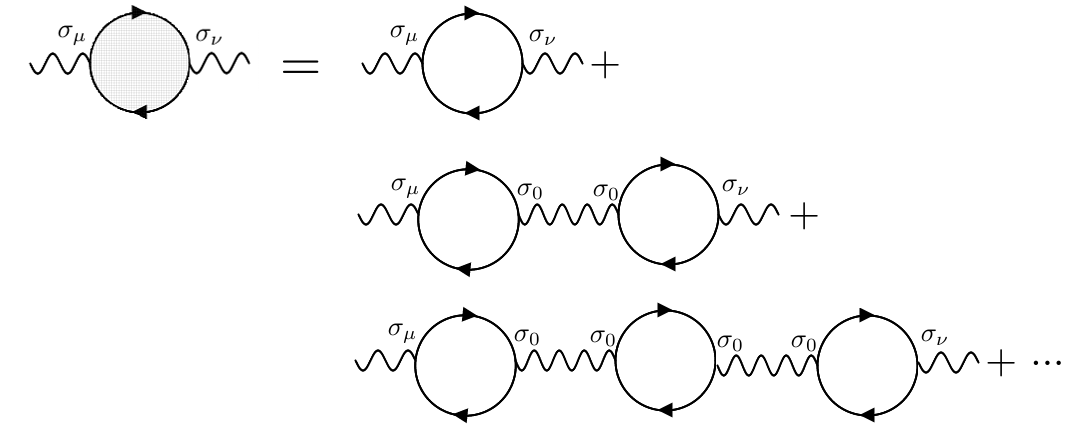}}
    \caption{Diagrammatic representation of Coulomb (density-density) interaction correction to current tensor correlation function. 
    Here $\sigma_{\mu(\nu)}$ represents the component of current vector and  $\sigma_0$ points to the zeroth component of current i.e. 
    the charge density operator. The Coulomb interaction only connects $\sigma_0$ vertices. }
    \label{RPA.fig}
\end{figure}

{\em Coulomb interaction correction:---}
In Dirac solids the continuum limit is described by the Minkowski spacetime, $\eta_{\mu\nu}$.
In such systems, there is no off-diagonal component on the first row or column of $\eta_{\mu\nu}$. 
In such systems, within the random phase approximation (RPA),
the Coulomb interaction being a density-density interaction will not be able to give
interaction induced corrections to the transverse component of the polarization. However, as we will show in this
section, with the metric $g_{\mu\nu}$ of Eq.~\eqref{tmetric.eqn}, due to the mixing of space and time (non-zero $g_{0i}$ components),
both longitudinal and transverse components of polarization will be corrected by the instantaneous Coulomb interactions. 

Let us use the covariant form of the tensor $\tilde \Pi$ and sum the RPA-like geometric series of Fig.~\ref{RPA.fig}. 
The Coulomb interaction operates in the density-density channel. The charge density is the $\mu=0$ component of the
current operator. Therefore both ends of the wiggly interaction lines in Fig.~\ref{RPA.fig} carry index $\mu=0$ only~\cite{VignaleBook,Chang2018}. 
Straightforward algebra for the above geometric series gives~\cite{Chang2018},
\be
\tilde{\Pi}^{\mu \nu}_{\rm RPA}(\Omega, {q})= \tilde\Pi^{\mu \nu}(\Omega, {q})+\frac{\tilde\Pi^{\mu 0}(\Omega, {q}) v(q) \tilde\Pi^{0 \nu}(\Omega, {q})}{1-v(q) \tilde \Pi^{00}(\Omega, {q})},
\label{RPAt.eqn}
\ee
 in which corrected current tensor elements is denoted by $\tilde{\Pi}_{\rm RPA}^{\mu \nu}(\Omega, {q})$. 
The "$00$" component of the above equation gives the celebrated RPA expression~\cite{VignaleBook}. 
In the case of Minkowski spacetime where metric is given by $\eta_{\mu\nu}$, the only source of 
non-zero contribution to "$\mu 0$" components in the second term of the above equation is proportional to $q^\mu q^0$ (see Eq.~\ref{covgraphene.eqn}).
Therefore the spatial component "$ij$" of the second term in Eq.~\eqref{RPAt.eqn} will be proportional to $(q^0)^2q^iq^j$. 
Now any tensor in space proportional to $q^iq^j$ is purely longitudinal. This is how in Minkowski spacetime, 
the Coulomb interaction can not generate RPA corrections in the transverse channel. But in the spacetime
of tilted Dirac fermions where the metric is $g_{\mu\nu}$ given in Eq.~\eqref{tmetric.eqn} the situation is different. 
In this case, the correction term will be given by,
\be
   \frac{v(\bsq)}{1-v(\bsq)\Pi^{00}}\left[(q^0)^2q^iq^j+q^4\zeta^i\zeta^j-q^2q^0(\zeta^iq^j+\zeta^j q^i) \right],\nn
\ee
which has both longitudinal and transverse components. We have defined $\zeta^i=\zeta_i$.
Note that $\bs{\zeta}$ is a parameter that defines the new geometry in Eq.~\eqref{tmetric.eqn} and 
is {\em not} a vector in this particular spacetime. 

In a standard Minkowski spacetime, the only way to generate interaction corrections in
transverse channels is to think of Thirring interactions~\cite{Thirring1958} of the form $(\bar\psi j^\mu\psi)(\bar\psi j_\mu\psi)$. 
The pure Coulomb forces (corresponding to $\mu=0$ component of the above Thirring interaction) as argued
above, will not be able to generate interaction induced corrections in the transverse channel. 
But the tilted Dirac solids furnish a unique solid state system where Coulomb interactions alone
are able to generate interaction induced corrections to transverse polarization. 

{\em Summary and outlook:---}
In this paper, we have calculated the polarization tensor for undoped tilted Dirac solids in 2+1 dimensions. 
The explicit calculation of the tensor shows that this tensor acquires a covariant representation~\eqref{tilted-cov.eqn},
where the metric tensor employed in this equation is given by Eq.~\eqref{tmetric.eqn}. 
This covariance of the representation~\eqref{tilted-cov.eqn} ensures that the same expression
holds in 3+1 dimensions for tilted Weyl semimetals as well. Furthermore, it satisfies the Ward identity. 
Therefore as far as the electromagnetic response of the system is concerned, the geometry of the spacetime
felt by electrons is given by metric~\eqref{tmetric.eqn}. The isometries of this metric will be different
from the Lorentz transformations of the Minkowski spacetime. Instead of Lorentz boosts, one will have a deformation of Lorentz boosts, etc~\cite{JafariEB}. 

When the Coulomb interactions are included, the deviation of metric $g_{\mu\nu}$ from the metric
$\eta_{\mu\nu}$ of the Minkowski spacetime combines the spatial and temporal components of the $j^\mu$
in the sense of Eq.~\eqref{tildehat.eqn}. This gives rise to interaction induced corrections in the
transverse part of the polarization tensor. This has no counterpart in ordinary electron "liquids"~\cite{Katsnelsonbook,VignaleBook}. 
In ordinary quantum liquids, including graphene, the collective charge density oscillations 
always correspond to longitudinal electric field oscillations. However, solids with tilted Dirac cone are an interesting form of electron liquids 
where collective charge oscillations can also be accompanied by transverse electric field propagation. 
This implies that in a tilted Dirac cone material the strength of the transverse photon is set by the electric
charge itself. This is in contrast to Minkowski spacetime where the Amperean forces (forces between two 
current carrying wires) arising from trasnverse photons are parametrically small by a factor of $(v/c)^2$~\cite{Kargarian2016Amperean}. 
Therefore the Amperean forces are parametrically a very small effect in Minkowski space. 
Amperean forces arising from emergent gauge field in strongly correlated systems can be a possible 
rout to non-Fermi liquid behavior~\cite{Senthil2007Amperean}. In this respect, the tilted Dirac cone
materials can be viewed as another platform where due to tilt parameter ${\bs \zeta}$, the Coulomb 
forces can be coupled to transverse photons. Given the low dimension of systems such as 
Borophene~\cite{Zabolotskiy2016,Zhou2014} and organic compounds~\cite{Katayama2006,Suzumura2014}, 
the Coulomb forces in these systems are likely to lead to non-Fermi liquid behavior as the 
transverse photons can not be screened. This might have relevance to anomalous Kondo effect in
organic materials that host 2+1 dimensional tilted Dirac cone~\cite{Suzumura2014,Tajima2002}. 

Combination of the above geometric description of the polarization tensor with axial term of $3+1$ dimensional 
Weyl systems and investigation of the fate of axial anomaly in deformed Minkowski spacetime~\eqref{tmetric.eqn}
is also an interesting question~\cite{Landsteiner2014,Kharzeev2016chiral,Beenakker2017Chiral}.
The fate of various collective excitations such as plasmons~\cite{Hwang2007}, triplons~\cite{Jafari2002,Jafari2005,Jafari2009,Jafari2012},
surface plasmon polaritons~\cite{Hofmann2016}, spin-plasmons~\cite{PoliniSpinPlasmons} etc a non-Minkowski background can teach us 
valuable lessons about the interplay of the spacetime geometry of solid state systems and the collective excitations. 
Investigation of the connection between this unique plasmonic feature
with the kink in the plasmon dispersion of tilted Dirac fermions~\cite{SaharTilt1,SaharTilt2} is an interesting question.

{\em Acknowledgements:---}
S.A.J. was supported by research deputy of Sharif University of Technology, grant no. G960214 and the
Iran Science Elites Federation. 

\appendix
\section{Alternative compact form for $\tilde F$}
\label{alternative.sec}
The current matrix elements $\tilde F$ of the tilted Dirac fermions can alternatively be
expressed in a compact form without invoking the corresponding $F$ factors of graphene. 
To this end we first represent the tilt in term of magnitude $\zeta$ and angular variable $\theta_t$ by 
$\zeta_{x}=\zeta \cos\theta_t$, $\zeta_{y}= \zeta \sin\theta_t$  and  substitute  $\theta_\bsk\rightarrow \theta_\bsk+\phi$ and in Eq.~\eqref{form factor.eqn}, 
\bearr
  &&\tilde F^{xx}_{s,s'}(\bsk,\bsk')=f_{s,s'}(\bsk,\bsq,\theta_t),\nn\\&&
  \tilde F^{yy}_{s,s'}(\bsk,\bsk')=f_{s,s'}(\bsk,q,\phi-\pi/2,\theta_t-\pi/2),
\nn\\&&
  \tilde F^{xy}_{s,s'}(\bsk,\bsk')=\tilde F^{yx}_{s,s'}(\bsk,\bsk')=\frac{1}{2}\big[f_{s,s'}(k,q,\phi-\pi/4,\theta_t-\pi/4)\nn\\&&-f_{s,s'}(k,q,\phi+\pi/4,\theta_t+\pi/4)\big],
\nn\\
  &&\tilde F^{x0}_{s,s'}(\bsk,\bsk')=\tilde F^{0x}_{s,s'}(\bsk,\bsk')=f'_{s,s'}(\bsk,\bsq,\theta_t),
 \nn\\&& 
   \tilde F^{y0}_{s,s'}(\bsk,\bsk')=\tilde F^{0y}_{s,s'}(\bsk,\bsk')=f'_{s,s'}(\bsk,q,\phi-\pi/2,\theta_t-\pi/2),\nn\\
\eearr
and 
\be
   F^{00}_{s,s'}(\bsk,\bsk')= \frac{1}{2}\bigg\{ 1+s s'\cos(\theta_\bsk-\theta_{\bsk'})\bigg\},
\ee
 where,
\bearr
&&f_{s,s'}(\bsk,\bsq,\theta_t)=
  \frac{1}{2}\bigg\{ \zeta^2 \cos^2\theta_t[1+s s'\cos(\theta_\bsk-\theta_{\bsk'})]\nn\\&&+[1+s s'\cos(\theta_\bsk+\theta_{\bsk'})]+  2\zeta \cos\theta_t [s \cos\theta_\bsk+s'\cos\theta_{\bsk'}] \bigg\},\nn\\
&&f'_{s,s'}(\bsk,\bsq,\theta_t)=
   \frac{1}{2}
   \bigg\{ \zeta \cos\theta_t[1+s s'\cos(\theta_\bsk-\theta_{\bsk'})]\nn\\&&+  [s \cos\theta_\bsk+s'\cos\theta_{\bsk'}] \bigg\}.\nn
   \eearr 
   
In undoped tilted Dirac cone with zero Fermi energy, at zero temperature the states with negative (positive) energy are occupied (unoccupied) which implies that the Fermi distribution function will be a step function, therefor the current response function Eq.\eqref{Pai.eqn} reduces to
\bearr
\tilde\Pi^{\mu\nu}(\boldsymbol{q},\Omega)=\frac{g_s }{A } \lim\limits_{\ep\rightarrow 0} && \sum_{k} \bigg\{ \frac{\tilde F^{\mu\nu}_{-,+}\left(\boldsymbol{k},\boldsymbol{q}\right)}{ \Omega-|\bsk|-|\bsk+\bsq|+i\ep}  -\nn\\&&\frac{\tilde F^{\mu\nu}_{+,-}\left(\boldsymbol{k},\boldsymbol{q}\right)}{ \Omega+|\bsk+\bsq|+|\bsk|+i\ep} \bigg\},
\label{PU.eqn}
 \eearr
By replacing $\bsk\leftrightarrow-(\bsk+\bsq)$ in the second part of Eq.~\eqref{PU.eqn}, the simplified representation is given by,
 \bearr
&&\tilde\Pi^{\mu\nu}(\boldsymbol{q},\omega)=\frac{g_s }{A} \lim\limits_{\ep\rightarrow 0} \sum_{k} \tilde F^{\mu\nu}_{-,+}\left(\boldsymbol{k},\boldsymbol{q}\right)\times \nn\\&&\bigg\{ \frac{1}{ \Omega-|\bsk|-|\bsk+\bsq|+i\ep}  -\frac{1}{ \Omega+|\bsk+\bsq|+|\bsk|+i\ep} \bigg\}.
 \label{paiu.eqn}
 \eearr
Again this expression is similar to the case of graphene with different $\tilde F$ functions and the replacement $\omega\to\Omega$.
This does not generate further difficulty, and the integrals can be evaluated using the same method employed in graphene~\cite{Wunsch2006,SaharTilt1}.
The result is identical to Eq.~\eqref{xx-yy-xy-undoped.eqn}.

\bibliography{mybib}

\end{document}